\documentclass[twocolumn]{aastex6}
\usepackage{natbib}
\usepackage{amsmath,amsthm,amssymb,amsfonts}

\shorttitle{Using Ice and Dust Lines to Constrain Disk Surface Density}
\shortauthors{Powell, Murray-Clay, \& Schlichting}

\begin{document}

\title{Using Ice and Dust Lines to Constrain the Surface Densities of Protoplanetary Disks}

\author{Diana Powell\altaffilmark{1,2}, Ruth Murray-Clay\altaffilmark{1}, and Hilke E. Schlichting\altaffilmark{3,4}}
\altaffiltext{1}{Department of Astronomy and Astrophysics, University of California, Santa Cruz, CA 95064}
\altaffiltext{2}{\href{mailto:dkpowell@ucsc.edu}{dkpowell@ucsc.edu} }
\altaffiltext{3}{Department of Earth, Planetary, and Space Sciences, University of California, Los Angeles, CA 90095}
\altaffiltext{4}{Department of Earth, Atmospheric and Planetary Sciences, Massachusetts Institute of Technology, 77 Massachusetts Avenue, Cambridge, MA 02139}

\begin{abstract}
We present a novel method for determining the surface density of protoplanetary disks through consideration of disk `dust lines' which indicate the observed disk radial scale at different observational wavelengths. This method relies on the assumption that the processes of particle growth and drift control the radial scale of the disk at late stages of disk evolution such that the lifetime of the disk is equal to both the drift timescale and growth timescale of the maximum particle size at a given dust line. We provide an initial proof of concept of our model through an application to the disk TW Hya and are able to estimate the disk dust-to-gas ratio, CO abundance, and accretion rate in addition to the total disk surface density. We find that our derived surface density profile and dust-to-gas ratio are consistent with the lower limits found through measurements of HD gas. The CO ice line also depends on surface density through grain adsorption rates and drift and we find that our theoretical CO ice line estimates have clear observational analogues. We further apply our model to a large parameter space of theoretical disks and find three observational diagnostics that may be used to test its validity. First we predict that the dust lines of disks other than TW Hya will be consistent with the normalized CO surface density profile shape for those disks. Second, surface density profiles that we derive from disk ice lines should match those derived from disk dust lines. Finally, we predict that disk dust and ice lines will scale oppositely, as a function of surface density, across a large sample of disks.
\end{abstract}

\section{Introduction}\label{intro}

Extrasolar planetary systems display a large diversity in both orbital architecture and the physical characteristics of the planets. This diversity could be the result of late stage planetary collisions \citep[e.g.][]{2016ApJ...817L..13I}, the properties and evolution of the initial gas disk \citep[e.g.][]{2016ApJ...825...29G}, different initial planetary formation locations in the disk \citep[e.g.][]{2015MNRAS.448.1751I} or a combination of these factors. The immediate initial conditions of planet formation are encapsulated in the protoplanetary disks that surround young stars. However, many disk characteristics remain largely unconstrained. Recent telescopic advances, particularly the Atacama Large Millimeter Array (ALMA), have enabled exploration of disks with unprecedented spatial resolution. These advances have already given us many insights into the spatial structure of disks \citep[e.g.][]{2016ApJ...819L...7N,2016Sci...353.1519P, 2016Natur.538..483T,2016ApJ...820L..40A, 2016ApJ...830...32W}. Here, we take advantage of this spatial resolution to propose new observational constraints on disk surface density, a property that is fundamental for understanding both disk evolution and planet formation.

Protoplanetary disk surface density cannot be measured directly because the majority of the disk mass resides in H$_2$ which is a symmetric particle that does not readily emit. Instead, densities have been inferred using observations of disk dust, CO, and HD. The reliability of each of these tracers has recently been called into question, leaving open the possibility that disk surface densities are entirely unconstrained \citep{1996ApJ...464L.169M, 2009ApJ...700.1502A,2010ApJ...723.1241A,2009ApJ...701..260I,2010ApJ...714.1746I, 2011A&A...529A.105G,2013Natur.493..644B,2014ApJ...788...59W, 2016ApJ...823...91S}. 

The first and most commonly used tracer of the surface density is the disk dust mass which is typically derived from resolved continuum observations or spectral energy distribution fitting \citep{2005ApJ...631.1134A,2002ApJ...568.1008C,2011A&A...529A.105G}. Using dust as a tracer of total disk surface density is fallible, however, as it requires an assumed dust-to-gas ratio. This ratio is not well constrained and can be altered from the ISM value of 10$^{-2}$ \citep{2011ARA&A..49...67W} through processes such as grain growth and particle drift \citep{2012ApJ...744..162A} and can further have a non-uniform value throughout a disk. 

CO gas has also been used as a tracer of the total gas present in disks \citep[e.g.][]{2012ApJ...757..129R}. However, recent observations have called into question the typically assumed abundance of CO \citep[10$^{-4}$ in warm molecular clouds;][]{1994ApJ...428L..69L}, suggesting that the existence of disk processes such as photodissociation may alter this value or that there may be a global depletion of gas phase carbon in disks such as TW Hya \citep{2001A&A...377..566V,2003A&A...402.1003D,2008A&A...488..565C,2016ApJ...823...91S}. In addition to this unknown, many CO lines are optically thick and are therefore unreliable measures of mass. The use of CO observations thus further requires careful consideration of lower optical depth CO isotopologues to estimate the gas mass to within an order of magnitude \citep{2014ApJ...788...59W, 2016ApJ...828...46A}.

More recently, observations of the HD $J=1-0$ line have been used to probe the gas mass in disks \citep{2013Natur.493..644B, 2016ApJ...831..167M}. HD is thought to be a good tracer of the total gas mass as the deuterium to hydrogen fraction is relatively well-known for objects near to the sun \citep{1998SSRv...84..285L}. However, because HD emits at temperatures above 20 K, HD emission will only trace the warm gas and thus provides a lower limit on the total gas mass present in a disk. 

Given the uncertainties that accompany these observational tracers, in this paper we choose to adopt an agnostic point of view in regards to surface density. We develop a novel method to derive this quantity through asserting that the dust line locations are determined by the microphysical process of particle drift. We use physics that has been studied extensively in previous disk models. Our contribution is to suggest a new interpretation of disk observations in light of this physics. 

Previous theoretical work indicates that, for the outer regions of evolved disks, drift dominates the processes of growth and collisional fragmentation in determining the maximum particle size at a particular radial scale. Particle growth can be limited by either a lack of total time (disk age), by fragmentation, or by drift. For typically turbulent disks, the disk lifetime allows plenty of time for particles to grow. The particle size at a given radial scale is limited, however, as particles are removed from the outer disk due to particle drift at a smaller maximum size than could be removed by collisional fragmentation. The maximum particle size at a given radius is thus defined as the size for which the growth timescale ($t_{\text{grow}}$) and the drift timescale ($t_{\text{drift}}$) are equal (i.e. $t_{\text{drift}} = t_{\text{grow}}$). This is described both numerically and analytically in \citet{2012A&A...539A.148B} and  \citet{2014ApJ...780..153B} as the ``late phase" of disk evolution in which dust growth has progressed such that it is limited by the removal of larger grains via the process of radial drift for roughly sub-centimeter sized particles. 

Thus, the maximum particle size at a given radial location is limited by particle drift which will remove all particles with a drift timescale less than the age of the system (i.e. $t_{\text{disk}}$). This indicates that the equilibrium particle size at a particular disk location is limited by the time in which that particle has been able to interact dynamically with the disk gas. The maximum disk radius where we would therefore expect to see emission from a given particle size is defined as the location for which the drift timescale (and thus the growth timescale) is equal to the age of the system (i.e. $t_{\text{drift}} = t_{\text{grow}} =  t_{\text{disk}}$). These locations can be seen observationally as the point where emission drops off at an observed wavelength ($\lambda_\text{obs}$) where we assume that the observed particle size is roughly equal to $\lambda_\text{obs}$. Any particle larger than this size would result in a shorter drift and/or growth timescale and would therefore reach an equilibrium location at a shorter radial scale. We note that we expect the equilibrium particle size at a particular radial location to decrease with increasing system age such that we expect larger disk radial scales at the same $\lambda_\text{obs}$ for younger disks, limited only by the size of the gas disk itself. 

In this paper we thus make the assumption that $t_{\text{drift}} = t_{\text{grow}} = t_{\text{disk}}$ to derive fundamental disk properties. We demonstrate that, if validated, this assertion allows us to recover the surface density distribution for observed disks. To test this central premise of our modeling we use two sets of observations that give empirical information about disk \textit{dust} and \textit{ice} lines.

The first set of observations is a collection of recent observations, using ALMA, the Jansky Very Large Array (JVLA) and the sub-millimeter array (SMA), of TW Hya that demonstrate that the disk radial scale is distinctly smaller at longer wavelengths \citep{2014A&A...564A..93M,2015ApJ...799..204C,2012ApJ...744..162A}. These observations provide information about the distribution of dust grains throughout the disk. We describe these observations in terms of disk \textit{dust lines} which refer to the disk radial scale as it corresponds to a particular wavelength. We use these dust lines to derive disk surface densities through equating $t_{\text{disk}}$ and $t_{\text{drift}}$ as described in Section \ref{drift}.

ALMA also provides direct observations of disk ice lines, either through direct measurements of CO gas line emission \citep{2016ApJ...819L...7N,2016ApJ...823...91S} or through indirect measurements of N$_2$H$^+$, which is only present in large abundance when CO freezes out \citep{2013Sci...341..630Q, 2015ApJ...813..128Q}. For TW Hya, the observation of N$_2$H$^+$ yields an ice line location of $\sim$ 30 AU \citep{2013Sci...341..630Q} while the emission from the C$^{18}$O $J=3-2$ line indicates an ice line of $\sim$ 10 AU \citep{2016ApJ...819L...7N}. The direct observations of CO emission also give insight into the surface density contributed by CO \citep{2015ApJ...799..204C,2012ApJ...757..129R}. The CO ice line location depends on the CO surface density through grain adsorption rates and particle drift. These observations therefore provide a constraint on the disk surface density and CO fraction.

After discussing representative parameters for our fiducial disk, TW Hya (Section \ref{TWH}), we explain how we derive disk surface densities from dust lines (Section \ref{dust}). In Section \ref{ice} we provide additional tests of our model through considering the disk ice lines. In Section \ref{tests} we present a description of three observational diagnostics of our model and an application to a larger range of disk parameter space. If these diagnostics confirm our interpretation, this will provide a new way to observationally measure disk masses and surface density profiles. We conclude with a paper summary and a discussion of the presented observational diagnostics in Section \ref{discussion}.

\section{Parameters for Fiducial Disk TW Hya}\label{TWH}

We adopt TW Hya as our fiducial protoplanetary disk because it is the nearest observed disk ($d = 54 \pm 6$ pc) that is nearly exactly face-on \citep[$i \sim 7^\circ$;][]{2004ApJ...616L..11Q} and hence boasts a wealth of observational data. TW Hya is a long-lived disk \citep[$t_\text{disk} = 3-10$ Myr;][]{2006A&A...459..511B,2011ApJ...732....8V} that is likely an unusually massive representative of a class of evolved protoplanetary disks as the typical disk lifetime is thought to be a few million years \citep{2009AIPC.1158....3M}. We note that disks are typically assumed to have the same ages as their host stars and estimates of stellar ages are subject to observational uncertainties. For our discussion of TW Hya we use an approximate age of 5 Myr, however, $t_\text{disk}$ could be treated as a tunable quantity as appropriate. 

We assume a temperature structure for TW Hya that is dominated by passive stellar irradiation. We expect that this model holds for the outer disk and note that TW Hya may be irradiation dominated at all but the very inner radii (\citet{2007A&A...473..457D} Figure 3, based on models from \citet{2006ApJ...638..314D}). Our parameterization of the disk midplane temperature follows \citet{1997ApJ...490..368C} where the canonical temperature profile is: 

\begin{equation}\label{canon} 
T(r) = T_0\times \left(\frac{r}{r_0}\right)^{-3/7}
\end{equation}

\noindent where the coefficient $T_0$ is a function of stellar luminosity and stellar mass, defined at $r_0 = $ 1 AU, and is determined via: 

\begin{equation}
T_0 = L_\star^{2/7}\left(\frac{1}{4\sigma_{SB}\pi}\right)^{2/7}\left(\frac{2}{7}\right)^{1/4}\left(\frac{k}{\mu G M_\star}\right)^{1/7}.
\end{equation}\label{T0}

We adopt the following parameters for TW Hya: $L_\star = 0.28 L_\odot$, $M_\star = 0.8 M_\odot$, and $\mu = 2.3\text{m}_\text{H}$ assuming a hydrogen/helium disk composition \citep{2013Sci...341..630Q,2007ApJ...660.1556R}. Using Equation \ref{T0}, we derive $T_0 \sim 82$ K. We note that our derived midplane temperature profile is in good agreement with the observationally constrained midplane temperature derived in \citet{2015ApJ...799..204C} as well as the upper limit on the midplane temperature from \citet{2016ApJ...823...91S}. We vary the normalization of this temperature profile in Section \ref{driftvclassic} and discuss the effect that this has on our ice line derivations. We also note that a factor of 2 change in temperature normalization ($T_0$) leads to a change of a factor of $\sim 0.7$ in our surface density profile derived in Section \ref{drift}.

Spatially resolved CO observations of TW Hya have been well fit by the following surface density profile:

\begin{equation}\label{sigma_0}
\Sigma(r) = \Sigma_c\left(\frac{r}{r_c}\right)^{-\gamma}\exp\left[-\left(\frac{r}{r_c}\right)^{2-\gamma}\right],
\end{equation}

\noindent which follows from the self-similar solution to the viscous evolution equations as shown in \citet{1974MNRAS.168..603L} and \citet{1998ApJ...495..385H}. This profile is a shallow power law at small radii and follows an exponential fall off at radii larger than the critical radius, $r_c$. Using an assumed CO abundance of $\sim 10^{-4}$ n$_\text{H}$ \citep[the standard CO fraction in warm molecular clouds;][]{1994ApJ...428L..69L}, \citet{2012ApJ...757..129R} derive best fit parameters for TW Hya of $r_c = 30$ AU, $\gamma =1$, and $\Sigma_c \sim 0.5$. 

As we move forward with our discussion of TW Hya we accept the best fit parameters for all values mentioned above, except for $\Sigma_c$, which relies on an assumed CO abundance. Instead, we treat $\Sigma_c$ as a free parameter. This is motivated by discrepancies between assumed and derived CO abundances in disks. We further note that there is also a potential discrepancy between the measured and derived mass accretion rates for TW Hya.

TW Hya has an average measured accretion rate of $\sim 1.5 \times 10^{-9}$ M$_\odot$ yr$^{-1}$ \citep{2012ApJ...760L..21B}. As a consistency check, we can use the surface density profile from Equation \ref{sigma_0} to derive an approximation of the mass accretion rate using the following expression \citep{2012MNRAS.419..925J}:

\begin{equation}\label{Mdot}
\dot{M} = \frac{M_\text{disk}}{t_\text{disk}}
\end{equation}

\noindent where $\dot{M}$ is the mass accretion rate, $t_\text{disk}$ is the age of the disk, and $M_\text{disk}$ is the disk mass which we take to be the mass of TW Hya interior to the critical radius of 30 AU. This estimate for $\dot{M}$ is a rough approximation for protoplanetary disks under the assumption that the primary mode of disk evolution is accretion. For instance, a rate that is higher than this derived  $\dot{M}$ would quickly evolve the disk past the current state and a lower $\dot{M}$ would indicate that a process other than accretion drove the disk to evolve into its current state. 

Thus, if the disk age is a proxy for evolution timescale, Equation \ref{Mdot} gives an accretion rate of $\sim 10^{-11}$ M$_\odot$ yr$^{-1}$ for TW Hya using the best fit parameters from \citet{2012ApJ...757..129R} -- a value that is 2 orders of magnitude smaller than the observational value. This value is inconsistent with observations. However, it is important to note that the accretion rates onto pre-main sequences stars are likely variable or episodic in nature \citep[e.g.][]{2001MNRAS.324..705A,2013ApJ...769...21S, 2016MNRAS.458.1466H}. For episodic accretion to explain this discrepancy, TW Hya would have to be currently undergoing an episode of high accretion -- a result that is unlikely given the smooth, axisymmetric nature of the disk and its observed central cavity.

We further note that photoevaporation can also remove mass in the outer disk. Since this process reduces a disk's accretion rate onto its star for a given disk mass, if important, it would make Equation \ref{Mdot} an upper limit for $\dot{M}$, making its agreement with observed accretion rates worse. The rate of photoevaporative mass loss for typical fluxes is $< 10^{-10}$ M$_\sun$ yr$^{-1}$ \citep{2006MNRAS.369..216A}, which is less than TW Hya's observed accretion rate, so it is likely subdominant. Thus, while the estimate of the mass accretion rate from Equation \ref{Mdot} is not necessarily conclusive, it nevertheless provides a reason to believe that mass in the disk may be higher than indicated by CO observations. 

TW Hya also has an observational lower limit of total gas mass of 0.05 M$_\sun$ from HD measurements of the warm gas in the disk \citep{2013Natur.493..644B}. The mass estimate is inconsistent with the mass estimate from the CO observations and gives an accretion rate of $\sim 6\times 10^{-9}$ M$_\odot$ yr$^{-1}$ which is more consistent with the observed rate. 

We consider the discrepancy between the measurements of the CO emission, HD gas emission, and observed accretion rate to be additional motivation for treating $\Sigma_c$ as a free parameter.

\section{Dust Lines}\label{dust}
The observed extent of TW Hya is wavelength dependent, ranging from a radius of $r\approx 25$ AU at a wavelength of $\lambda = 9$ mm to $r\approx 130$ AU at a wavelength of 1.6 $\mu$m as shown in Figure \ref{particlesize}. In particular, observations at 0.87 mm show a disk size of approximately 60 AU \citep{2012ApJ...744..162A,2016ApJ...820L..40A}, at 1.3 mm the disk size is around 50 AU \citep{2015ApJ...799..204C}, and at 9 mm the disk size is approximately 25 AU \citep{2014A&A...564A..93M}. We take the observed disk size of approximately 130 AU at 1.6 $\mu$m to indicate the total radial extent of the disk as this distance matches the observed radial extent of the CO emission \citep{2013Sci...341..630Q}. 

Recent observational work has found that the continuum emission at each wavelength exhibits a markedly sharp decrease over a very narrow radial range such that $\Delta r/r \lesssim 0.1$ \citep{2012ApJ...744..162A,2013A&A...557A.133D}. Thus, while these are only approximate radial sizes for the TW Hya disk, they are adequate for our preliminary physical interpretation. In our further discussion of TW Hya we allow the radius error bars to vary by $\pm 10$ AU and find little change in our theoretical modeling. These error bars are slightly larger than those derived from treating the systematic uncertainty in distance to the TW Hya system alone \citep[$d = 54 \pm 6$ pc or $d = 51 \pm 6$ pc][which gives uncertainties that roughly range from $\pm$ 3 AU at $\lambda_{obs} = 9$ mm to $\pm$ 7 AU at $\lambda_{obs} = .87$mm]{2007ASSL..350.....V,2005ApJ...634.1385M}. We inflate the error bars to account for error in measuring the disk radial scale without modeling the disk visibilities. We will improve on these error estimates in future work.

\begin{figure}
\epsscale{1.1}
\plotone{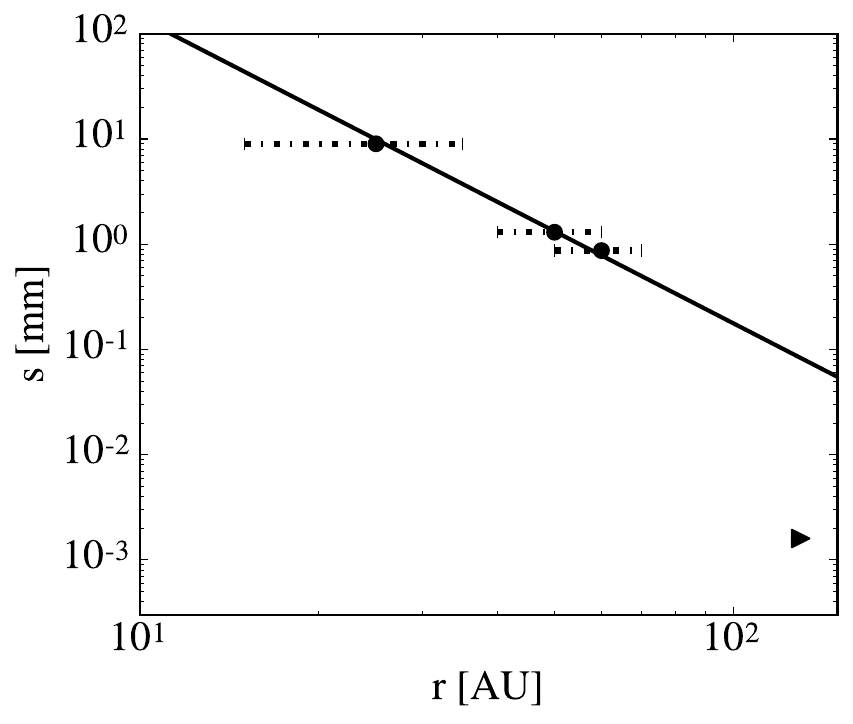}
\caption{The dominant particle size in the disk as a function of radius which follows a power law relationship. The three black points represent disk sizes derived from observations by \citet{2014A&A...564A..93M}, \citet{2015ApJ...799..204C}, and \citet{2012ApJ...744..162A,2016ApJ...820L..40A}. The upper limit arrow represents measurements of the radial extent of both the CO gas disk and the disk size of the smallest grains which we take to be indicative of the total radial extent of the disk \citep{2013ApJ...771...45D}. This is a theoretical lower limit as we expect that small dust sizes would have a larger radial extent if the gas disk was larger. The error bars shown correspond to our chosen nominal error for the disk radius of $\pm 10$ AU, consistent with the observationally sharp cut-off in disk emission (see text).}
\end{figure}\label{particlesize}

Models of disk emission from particles with a range of sizes indicate that $\lambda_{\text{obs}}$ roughly corresponds to the primary particle size, $s$, contributing to the observed emission \citep{2014ApJ...780..153B}. We therefore set the particle radius $s=\lambda_{\text{obs}}$ for all that follows and leave a more detailed evaluation of the particle size distributions for future work. We call the cutoff distance for emission at $\lambda_{\text{obs}} = s$ the ``dust line" for particles of size $s$. 

In Figure \ref{particlesize} we note that the observed locations of dust lines as a function of particle size (equivalent to the disk size as a function of $\lambda_{\text{obs}}$) can be well fit by a power law, with the exclusion of the disk size at the shortest $\lambda_{\text{obs}} = $ 1.6 $\mu$m. Because the radial extent of the 1.6 $\mu$m grains matches that of the CO gas, they are limited by the total disk size and not by the equilibrium that governs disk sizes at longer $\lambda_{\text{obs}}$. We conclude that this point acts as a theoretical lower limit for the radius at which you would expect to find these grains \citep{2012ApJ...757..129R,2013ApJ...771...45D}. Given their present location, they have not yet had to time to grow or drift due to the long dynamical timescales for micron sized particles in the outer disk.

The presence of dust lines in TW Hya indicate that a physical process is removing particles larger than a set maximum size from the outer disk. Particle growth can be truncated by particle fragmentation, particle drift, or a lack of total dynamical time. If we consider the case of TW Hya, the age of the system ($t_{\text{disk}}$) is sufficiently large that there is plenty of time for particle growth to occur. 

Previous work has shown that for the outer regions of evolved disks, such as TW Hya, particle drift is significant for sub-centimeter sized bodies such that these particles are not able to grow to a size that can be disrupted due to collisional fragmentation \citep[see][]{2012A&A...539A.148B,2014ApJ...780..153B}. This is physically intuitive as small particles in the outer disk drift faster as they grow larger (see Section \ref{drift}). 

This drift limited regime reaches an equilibrium such that the radial drift of particles imposes a size limit as large particles are removed faster than they can be replenished due to particle growth. Particles in the outer disk are also removed at a size that is smaller than the limit imposed by collisional fragmentation, for disks of standard turbulence, as the relative turbulent velocities between particles of these sizes is not sufficiently large for efficient destruction \citep{2012A&A...539A.148B}. Thus, the maximum particle size at a given radial location in the disk is given by the size for which the growth timescale is equal to the drift timescale (i.e. $t_{\text{grow}} = t_{\text{drift}}$). 

Smaller particles have longer drift timescales (see Section \ref{drift}) which means that the dust lines in a disk will evolve with disk age such that we expect the radial location of a dust line at a specific $\lambda_\text{obs}$ to decrease with increasing disk age. This therefore indicates that the dust line of a disk is determined by the time in which particles are able to drift, which is necessarily the age of the system. We can thus expect the fall-off of emission from a given particle size to correspond to the location where a particle's drift timescale is equal to the lifetime of the disk ($t_{\text{drift}} = t_{\text{disk}}$). Therefore, the system can be described at any given time by an equilibrium state in which $t_{\text{drift}} = t_{\text{grow}} = t_{\text{disk}}$. This new picture of disks is summarized in Figure \ref{disk_cartoon}.

\begin{figure}
\epsscale{1.2}
\plotone{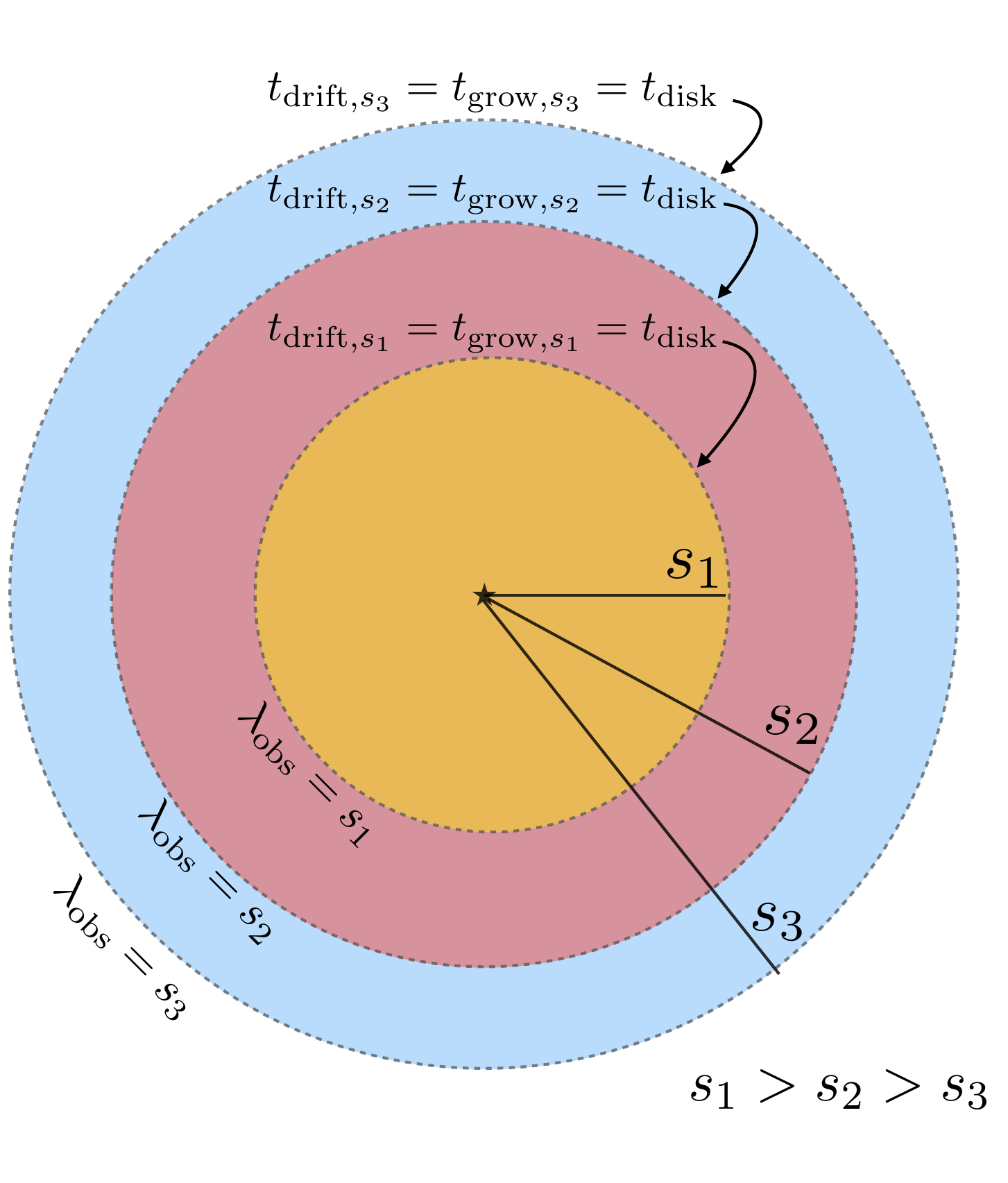}
\caption{Cartoon of our model for disk dust lines (dashed lines). Particle sizes are denoted $s_1$, $s_2$, and $s_3$ where $s_1 > s_2 > s_3$. Particles of size $s_1$ are present in the disk throughout the yellow region. Particles of size $s_2$ extend throughout the yellow and red regions while particles of size $s_3$ are present throughout all depicted disk regions. At the dust lines for each particle size the growth and drift timescale are equal to the age of the system. When observed at $\lambda_\text{obs} = s_1$ only the yellow region of the disk will be observed, while for $\lambda_\text{obs} = s_2$ the disk extends radially to the end of the red region and for $\lambda_\text{obs} = s_3$ the disk appears to extend to the end of the blue region.}
\end{figure}\label{disk_cartoon}

Using the dust line observations of TW Hya in conjunction with our theoretical premises we can now consider the dominant physical processes of growth and drift and use these calculations to derive the total disk surface density for TW Hya as well as the dust-to-gas ratio in the outer disk. We do this through the use of our central modeling premise that $t_{\text{drift}} = t_{\text{grow}} = t_{\text{disk}}$.

\subsection{Radial Drift}\label{drift}

As discussed above, we assume that the disk radial scale as a function of wavelength is set by the distance that a particle of radius, $s = \lambda_{\text{obs}}$, can drift in the age of the system. We therefore set the drift timescale $t_{\text{drift}}$ equal to the lifetime of the system $t_{\text{disk}}$.  

In a protoplanetary disk particle drift occurs because the gas disk orbits at a sub-Keplerian velocity due to an outward pressure gradient \citep{1977MNRAS.180...57W}. The particles in the disk continue to rotate at a Keplerian velocity ($v_k \equiv \Omega_k r$) and experience a headwind from the gas. This headwind causes the particles to loose angular momentum and drift radially inwards \citep{1977MNRAS.180...57W,2002ApJ...581.1344T}. The amount of drift that a particle experiences depends on how well-coupled the particle is to the gas, quantified by a dimensionless stopping time: $\tau_s \equiv \Omega_k t_s$ where $\Omega$ is the Keplerian frequency and

\begin{equation}\label{stopping}
\begin{array}{@{} r @{} c @{} l @{} }
&t_s &{}=\displaystyle
\begin{cases}
\rho_s s/\rho c_s & s< 9\lambda/4, \text{  Epstein drag},\\
4\rho_s s^2/9 \rho c_s \lambda &  s> 9\lambda/4, \text{Re} \lesssim 1 \text{  Stokes drag}
\end{cases}
\end{array}
\end{equation}

\noindent (summarized in \citet{2010AREPS..38..493C}). Here $\rho$ is the gas midplane density, $\rho_s = 2$ g cm$^{-3}$ is the density of a solid particle, $s$ is the particle size, and $\lambda = \mu/\rho \sigma_\text{coll}$ is the gas mean free path where $ \sigma_\text{coll} = 10^{-15}$ cm$^2$. 

The radial particle drift velocity is:

\begin{equation}\label{rdot}
\dot{r} \approx -2\eta \Omega r \left(\frac{\tau_s}{1+\tau_s^2}\right)
\end{equation}

\noindent where $\eta \approx c_s^2/2v_k^2$, $c_s$ is the sound speed of the gas, and $v_k$ is the Keplerian velocity \citep[see the review by][]{2010AREPS..38..493C}. 

We can now derive an equation for drift timescale ($t_{\text{drift}} = \left|r/\dot{r}\right|$) that directly depends on the disk surface density, $r$, and $s$. We first note that the particle sizes we consider ($s < 1$ cm; see Section \ref{dust}) typically interact with the disk gas via the Epstein law for gas drag and have a dimensionless stopping time less than 1. Once our calculation is complete we verify that the Epstein regime applies for our reconstructed surface density profile and that $\tau_s <1$. 

We approximate a dimensionless stopping time in the Epstein regime as $\tau_s = \Omega\rho_ss/\rho c_s$. We can rewrite this quantity by noting that the disk's scale height $H = c_s/\Omega$ and its surface density in gas $\Sigma_g = H\rho$ to obtain $\tau_s = \rho_s s/\Sigma_g$. As found in \citet{2013ApJ...775..136R}, increasing temperature with height can inflate the aspect ratio of an observed disk when layers several scale-heights above the midplane are probed. As most of the disk mass is contained within the scale height closest to the midplane, however, the temperature at the midplane sets the volumetric density in the disk to an order of magnitude as described above.

We reformulate the drift timescale through defining the parameter $v_0$ which physically corresponds to the maximum drift velocity. For a passively irradiated disk, $v_0 \equiv \eta v_k = c_s^2/2v_k$ varies very weakly with radius. We find that $c_s \propto r^{-3/14}$ where $c_s = \sqrt{kT/\mu}$ using Equation \ref{canon} for temperature as a function of radius. For the Keplerian velocity we find that $v_k \propto r^{-1/2}$ using Kepler's third law. This gives a parameter $v_0$ that is weakly dependent on radius: $v_0 \propto r^{1/14}$.

We can now write the drift timescale directly in terms of the surface density, radius, particle size, and the maximum drift velocity $v_0$:

\begin{equation}
t_{\text{drift}} = \left|\frac{r}{\dot{r}}\right| = \Omega^{-1}\left(\frac{r}{H}\right)^2\tau_s^{-1}\approx \frac{\Sigma r}{v_0 \rho_s s}
\end{equation}

We can now set the drift timescale equal to the age of the disk ($t_{\text{drift}} = t_{\text{disk}}$) and solve for the gas surface density at a given dust line radius. We thus derive the following equation for disk surface density as a function of radius. 

\begin{equation}\label{sigma}
\Sigma(r) =\frac{ t_{\text{disk}}v_0\rho_s s}{r}
\end{equation}

From this equation we see that $\Sigma(r) \propto s/r$ with $s(r)$ plotted in Figure \ref{particlesize}. This reformulation is thus particularly powerful as it gives a direct scaling between surface density and radius with only a very weak dependence on the temperature profile of the disk via the term $v_0$.

We now apply Equation \ref{sigma} to TW Hya again assuming that the dominant particle size $s$ is given by the wavelength of the observation ($\lambda_{\text{obs}}$). The resulting surface density points are well described by a steep power law function of approximately $r^{-4}$. Initially, this appears to be an unrealistically steep relation for the disk surface density as typical scalings for the minimum mass solar nebula are $\Sigma \propto r^{-1}$. However, as these points fall close to the observationally derived exponential fall-off range for the measurements of CO in \citet{2012ApJ...757..129R}, the data match this profile quite well when we allow $\Sigma_c$ to be a floating normalization factor (see Section \ref{TWH}). We find that a surface density normalization of $\Sigma_c \approx 10^{2.5}$, approximately 3 orders of magnitude larger than $\Sigma_c$ from Rosenfeld et al., adequately matches our derived surface density points. The derived surface density profile with error estimates is shown in Figure \ref{radius_errors}. We find that the particle sizes we consider in the outer disk are indeed still well within the Epstein regime as was assumed in our derivation of the disk surface density profile. 

We note that this surface density estimate depends on the age of the disk which may be uncertain by as much as a factor of 2 \citep[e.g.][]{2012ApJ...746..154P}. For TW Hydra the surface density normalization, $\Sigma_c$, ranges from $\sim$ 200 at 3 Myr to $\sim$ 800 at 10 Myr. We note that the upper end of this age range will lead to a derived disk surface density that is Toomre-Q unstable, however, due to the order of magnitude nature of this derivation those numbers could be revised through a more detailed calculation. This trend generally holds such that the derived disk surface density increases with an increase in estimated disk age.

\begin{figure}
\epsscale{1.1}
\plotone{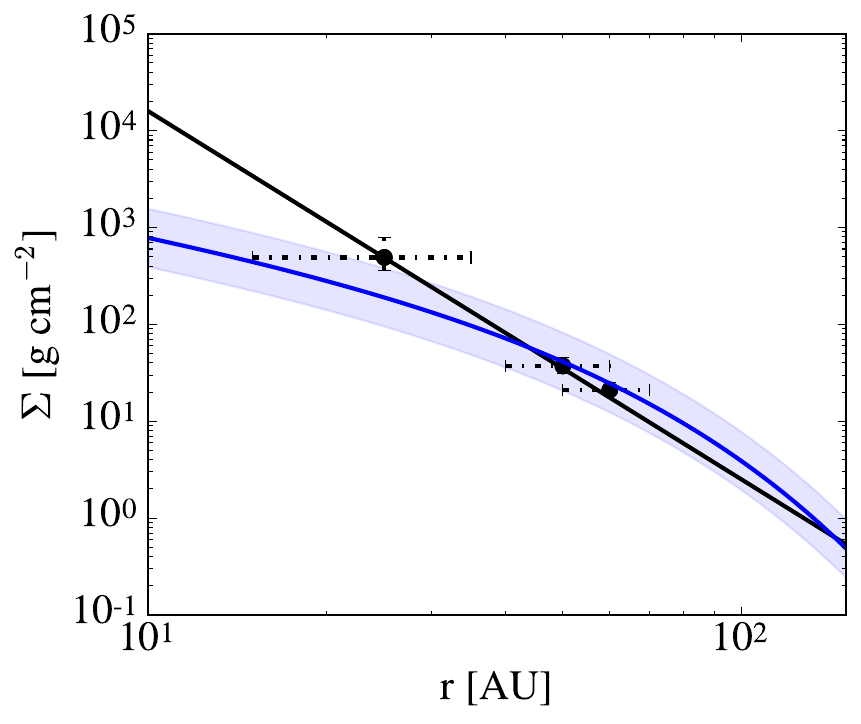}
\caption{Surface density of TW Hya (points) derived from Equation \ref{sigma} using the three observed disk sizes $r(s)$ with radius error bars of $\pm$ 10 AU and their corresponding errors in surface density. The surface density profile at these radii are well fit by an $r^{-4}$ power law shown by the solid black line. The normalized surface density profile is shown in blue with the corresponding shading region indicating the roughly normalized surface density profile with the inclusion of the radial error estimates. The normalized surface density profile is an $r^{-1}$ power law interior to the critical radius ($r_c$) of 30 AU and is then described by an exponential fall off at radii larger than $r_c$. We find that a surface density normalization of $\Sigma_c \approx 10^{2.5}$ adequately matches our derived surface density points.}
\end{figure}\label{radius_errors}

We can now convert our surface density profile to a derived gas mass as well as a mass accretion rate for TW Hya. We find that our derived gas mass of approximately 0.05 $M_\odot$ is consistent with the observational lower limit of 0.05 M$_\odot$ as derived from HD measurements of the warm gas \citep{2013Natur.493..644B}. We remark that this agreement implies no low-density cool gas and note that there is room within the errors for there to be a comparable amount of cold and hot mass present within the disk, however, to avoid being Toomre-Q unstable the mass cannot increase by a substantial amount.

Following the discussion in Section \ref{TWH}, with the inclusion of rough radius error estimates, we find a mass accretion rate of $\dot{M} \sim 4\times 10^{-9} - 2\times 10^{-8}$ M$_\odot$ yr$^{-1}$. Compared to the measured value of $\sim 1.5 \times 10^{-9}$ M$_\odot$ yr$^{-1}$ \citep{2012ApJ...760L..21B} our derived accretion rates give a larger value more in line with measured accretion rates for younger systems than the derived disk accretion rate using CO surface density observations as in \citet{2012ApJ...757..129R}. We note that TW Hya has an inner disk gap at 1 AU \citep{2016ApJ...820L..40A}. It is possible that this gap was formed by the presence of an accreting protoplanet which could be allowing only $\sim$ 10\% of the accretion flux onto the star \citep{2005A&A...433..247P,2007MNRAS.378..369N,2015ApJ...799...16Z,2015ApJ...803L...4E}. 

We also note that, given our estimated disk age, our derived surface density profile is Toomre-Q unstable for the higher region of our estimated errors. We note, however, that assuming a shorter age of the system ($\sim 3$ Myr) gives a stable disk surface density profile for all estimated errors. 

\subsection{Particle Growth and the Dust-to-Gas Ratio}

Given that TW Hya is an evolved disk in dynamical equilibrium, we can use an approximation of the growth timescale to determine the dust-to-gas ratio in the outer disk through allowing $t_{\text{grow}} = t_{\text{disk}}$ as discussed in Section \ref{dust}. We derive our growth timescale by first considering the growth rate in terms of particle size ($s$):

\begin{equation}\label{growthrate}
\dot{m} = \rho_{d} \sigma \Delta v
\end{equation}

\noindent where $\rho_p$ is the density of the particles which we take to be $\sim \rho_gf_d$ where $f_d$ is the dust-to-gas ratio and $\sigma = \pi s^2$ where $s$ is the size of the dominant particle at that radius $r$. In the outer disk we assume that the relative particle velocities $\Delta v$ are due to turbulence which we approximate as:

\begin{equation}
\Delta v = \alpha c_s^2\frac{t_L}{t_\eta}(\tau_{s,1}-\tau_{s,2})^2
\end{equation}

\noindent \citep{2007A&A...466..413O} where $t_L$ is the overturn time of the largest eddies which we take to be the orbital period, $t_\eta = Re^{-1/2}t_L$, $\tau_{s,1}$ is the stopping time of our dominant particle size, $\tau_{s,2}$ is the stopping time of the particle size that contributes to growth, and $\alpha$ is the parameter of ignorance for turbulence in a disk which we take to be a standard value of $10^{-3}$ \citep[e.g. ][]{2017ApJ...837..163R}. This $\Delta v$ is valid for particles that are tightly coupled to the gas. We note that direct observations of disk turbulence \citep{2011ApJ...727...85H,2016A&A...592A..49T} do not yet probe the disk midplane.

We are now able to solve for the dust-to-gas ratio in the disk through setting the growth timescale ($t_{\text{grow}}$) equal to the age of the system ($t_{\text{disk}}$) and solving for $f_d$. We do this while assuming that the particles grow through perfectly efficient sticking with particles that are similar in size. We make this approximation as most of the dust surface density will be in the largest grains such that these particles dominate particle growth. Given this assumption, typical values of $\Delta v$ for our TW Hya calculation are $\sim 3 - 10$ cm s$^{-1}$. For TW Hya this exercise gives an average dust-to-gas ratio of $f_d \approx 10^{-3}$ in the outer disk where we observe the disk dust lines. 

As a consistency check we can compare this derived dust-to-gas ratio with the observed dust surface density profile described in \citet{2012ApJ...744..162A}. We note that the total dust mass and the dust surface density profile are model dependent quantities with values that vary across the literature \citep{2002ApJ...568.1008C,2005ApJ...631.1134A,2008ApJ...678.1119H,2011ApJ...740...38H,2013A&A...559A..24K,2016A&A...586A..99H}. We thus choose the \citet{2012ApJ...744..162A} model as it is based off of 870 $\mu$m emission which is within the range of our dust line observations.

In their model, the dust surface density profile is well described by a shallow power law $\Sigma_d \propto r^{-0.75}$ until the emission falls off at roughly 60 AU. This profile is normalized such that $\Sigma_{d} = 0.39$ g cm$^{-2}$ at 10 AU. When we divide this dust surface density profile by our derived total surface density profile we find an average dust-to-gas ratio of $\sim 10^{-2.5}$ which is roughly consistent with our derived dust-to-gas ratio above. 

We further compare our dust surface density points at the three dust line locations using our derived dust-to-gas ratios with the best fit model from \citet{2012ApJ...744..162A} in Figure \ref{sigmadust}. We find a dust surface density that is systematically lower than the best-fit model from the 870 $\mu$m emission, however, without error bars on this observationally derived model it is difficult to know if our results are discrepant. We determine an extremely rough estimate of the uncertainty of this model by propagating uncertainties on the mass opacity, flux calibration ($\sim 10$ \% systematic uncertainty), and distance (a few percent uncertainty). The largest source of uncertainty in the \citet{2012ApJ...744..162A} model is the mass opacity. We adopt an uncertainty on this value of roughly $\sim 99$ \% as the range in assumed mass opacities for dust at $\sim$ 870 $\mu$m ranges from $\sim$ 10$^{-2} - 10^0$ in the relevant literature \citep[e.g.][]{1983QJRAS..24..267H,1994ApJ...421..615P,1996A&A...311..291H,2005ApJ...631.1134A,2012ApJ...744..162A}. The error on the mass opacity thus dominates the uncertainty in the dust measurement. We note that our points represent the dust surface density comprised of the large particles in some set size distribution and that including the contribution of smaller grains to the total surface density should increase our answer by a few tens of percent. 

We also compare to dust depletion in a time evolving disk as shown in \citet{2014ApJ...780..153B} (see their Figure 4a) where they find a solid depletion of up to two orders of magnitude in their fiducial model for the outer regions of late stage disks. We therefore find that our estimate of dust depletion is consistent with their model results for evolved disks. 

\begin{figure}
\epsscale{1.1}
\plotone{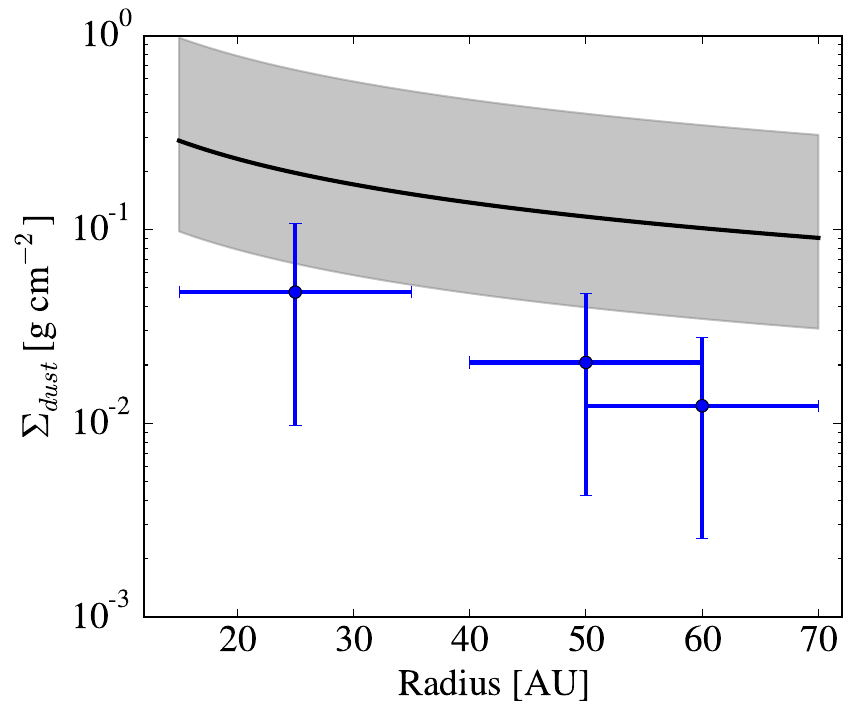}
\caption{The dust surface density of TW Hya (points) determined from our derived dust-to-gas ratio in conjunction with our dust line derived total surface density profile. These points can be compared to the best-fit dust surface density profile (black line) from \citet{2012ApJ...744..162A} for TW Hya derived using 870 $\mu$m emission. We find a dust surface density that is systematically lower than the best-fit model from the 870 $\mu$m emission. We provide a rough estimate of the uncertainty surrounding the dust surface density profile from \citet{2012ApJ...744..162A} (gray shaded region, see text) and find that our results are not discrepant within the assumed error.}
\end{figure}\label{sigmadust}

In conclusion, by considering particle drift we derive a total disk surface density for TW Hya that is roughly 3 orders of magnitude larger than the surface density derived from CO observations through considering particle drift. We further derive a dust-to-gas ratio of approximately $10^{-3}$ in the outer disk ($\sim 25-60$ AU).

\section{Ice Lines}\label{ice}

We can now further our discussion of the disk surface density through a careful treatment of the molecular ice lines. The three ice lines that are most frequently calculated are H$_2$O, CO$_2$, and CO as these species are considered to be in relatively high abundance \citep{2011ApJ...743L..16O}. To date, the CO ice line is the most readily observed due to its large radial distance from the central star \citep{2013Sci...341..630Q}. We thus primarily focus on the CO ice line in our calculations. However, when we expand our discussion to consider a larger parameter space of disk parameters, we further extend our arguments to the other volatile species as well. 

While we consider a passively heated disk (see Section 2), we note that the water ice line may be impacted by accretion heating. We include some plots of the H$_2$O ice line for reference, but these should be viewed with caution.

Our discussion focuses on the \textit{midplane} ice lines as opposed to the surface ice lines frequently discussed in the literature \citep{2016ApJ...818...22B,2016ApJ...823...91S}.  As a result we will compare our derived ice lines to observations that probe the midplane ice line directly \citep[via the C$^{18}$O line;][]{2015ApJ...813..128Q,2003A&A...399..773D} or indirectly \citep[via the N$_2$H$^+$ ion;][]{2013Sci...341..630Q,2015ApJ...813..128Q}. 

There are three pieces of physics that we consider in our ice line calculations: particle adsorption, desorption, and drift.

\subsection{Volatile Adsorption and Desorption}\label{classical}
The classic ice line calculation balances adsorption and desorption flux onto a grain to determine the ice line radius \citep{2009ApJ...690.1497H,2011ApJ...743L..16O}. We refer to this ice line as the `classical ice line'. Following Hollenbach et al. 2009, these two fluxes are quantified as:

\begin{equation}\label{ad}
F_{\text{adsorb}} \sim n_ic_s
\end{equation}

\begin{equation}\label{de}
F_{\text{desorb}} \sim N_{s,i}\nu_{\text{vib}}\text{e}^{-E_i/kT_{\text{grain}}}f_{s,i}
\end{equation}

\noindent where $n_i$ is the relevant gas density species, $c_s$ is the sound speed, $N_{s,i}\approx 10^{15}$ sites cm$^{-2}$ is the number of adsorption sites per volatile per cm$^2$, $\nu_{\text{vib}} = 1.6\times10^{11}\sqrt{(E_i/\mu_i)}$ s$^{-1}$ is the molecules vibrational frequency in the surface potential well, $E_i$ is the adsorption binding energy in units of Kelvin, and $f_{s,i}$ is the fraction of the surface adsorption sites that are occupied by species $i$ (which we take to be unity). Finally, we assume that $T_{\text{grain}} = T$, meaning that the dust and gas have the same temperature in the disk mid-plane. 

Balancing Equations \ref{ad} and \ref{de} allows for us to solve for the freezing temperature of a species as a function of radius for a given disk surface density profile. We then locate the classical ice line by finding the disk radius where the molecular freezing temperature is equal to the disk temperature. This self-consistent method allows us to determine how the ice line location changes as a function of both disk surface density and temperature. 

\subsection{The Influence of Particle Drift}\label{driftice}
Particle drift, as described in Section 3, influences the location of the ice lines \citep{2015ApJ...815..109P,2016ApJ...833..203P}. This is because particles that drift faster can cross the ice line before desorbing, thus potentially moving the location of the ice line inwards. The drift ice line location can be calculated by setting the desorption timescale, 

\begin{equation}\label{desd}
t_{\text{des}} = \frac{\rho_s}{3\mu_i m_\text{H}}\frac{s}{N_{s,i}\nu_{\text{vib}}\text{e}^{-E_i/kT_{\text{grain}}}},
\end{equation}

\noindent (where $\mu_i$ is the molecular weight of the desorbing species) equal to the drift timescale (the radial location of a particle divided by Equation \ref{rdot}) and solving for the desorption distance, $r_{\text{des}}$ \citep[as verified by time evolving calculations in][]{2015ApJ...815..109P}. This is done analytically for the small stopping time approximation in the Appendix of \citet{2015ApJ...815..109P}. We extend this calculation to the $\tau_s > 1$ regime through making no approximations in Equation \ref{rdot} in regards to $\tau_s$. Instead, we use the two distinct stopping time expressions given both Stokes and Epstein regimes (Equation \ref{stopping}). We use the SciPy subroutine fsolve \citep{Jones} to solve for radius through balancing the drift timescale with the desoprtion timescale ($t_\text{drift} = t_\text{des}$) where these timescales are given by $t_\text{drift} = \left|r/\dot{r}\right|$, where $\dot{r}$ is given by Equation \ref{rdot}, and Equation \ref{desd}. We consider the radial dependence of each term without approximation, which allows us to derive a completely self-consistent estimate of $r_{\text{des}}$ for any arbitrary surface density and/or temperature profile.

\begin{figure*}
\epsscale{1.}
\plottwo{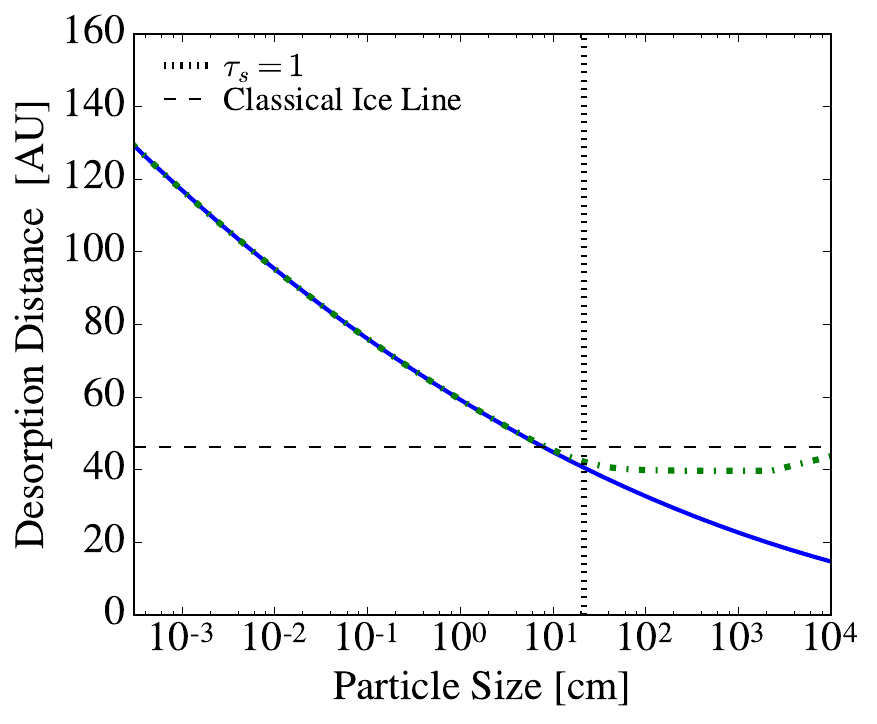}{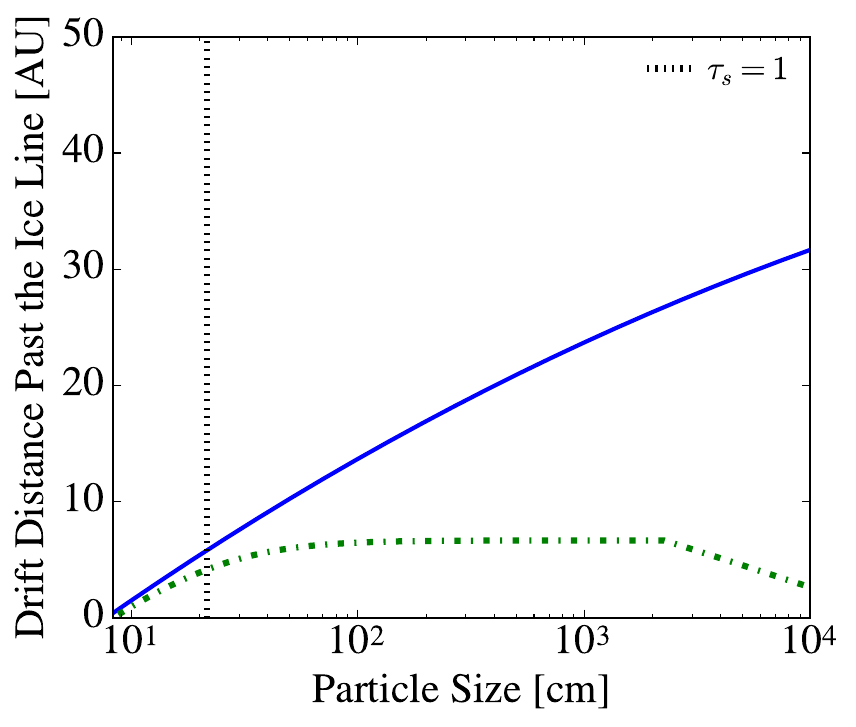}
\caption{Left: The desorption distance, the location where the drift timescale equals the desorption timescale, calculated using the analytic method from  \citet{2015ApJ...815..109P} (blue, solid) and our new extended solver (green, dashed). We find great agreement for particles with a stopping time less than unity. Right: The distance that a CO particle is able to drift past the classical CO ice line before desorbing using the analytic method from  \citet{2015ApJ...815..109P} (blue, solid) and our new self-consistent solver (green, dashed). Using our extended solver we see that, for this comparison case, particles smaller than $\sim$ 10 cm do not drift past the ice line and are not shown. Particles larger than $\sim$ 8 cm do experience drift in this case, with the maximum drift reached at a particle size slightly larger than $\tau_s \approx 1$ that reaches a near constant value at larger stopping times.}
\end{figure*}\label{drift_distance}

We compare the difference between the drift ice line and the classical ice line for our new self-consistent solver and the analytic solver from \citet{2015ApJ...815..109P} in Figure \ref{drift_distance}.  We use the disk temperature and surface density profiles from \citet{2015ApJ...815..109P} and find that particles smaller than $\sim$ 10 cm ($\tau_s \sim 0.5$) do not drift past the CO ice line. Particles larger than $\sim$ 10 cm do experience drift, with the maximum drift reached at a $\tau_s \approx 1$ that then reaches a near constant value at higher stopping time. We note that, while small particles do not drift to the same radial location as large particles, accretion of disk gas has been shown to pull smaller particles in past the ice line as shown in \citet{2015ApJ...815..109P}.

The nearly constant drift distance at particle sizes with a stopping time greater than 1 occurs due to the interplay between two different pieces of physics. The first is that the larger the particle size the longer it takes for that particle to desorb and the further it can drift past the ice line. The second is that the maximum drift velocity occurs for particles with $\tau_s = 1$. The decrease in drift velocity past a stopping time of one is offset by the increase in particle size until the regime changes from Epstein to Stokes. The drift timescale for a particle in the Epstein with $\tau_s >> 1$ is $t_\text{drift} =\left|r/\dot{r}\right| \sim \tau_s/\eta \Omega \sim \rho_s s/\rho c_s\eta$. The desorption timescale is simply proportional to the particle size ($t_\text{des} \propto s$) with a desorption constant such that $t_\text{des} \sim C_\text{des}s$. When we equate these two timescales we find that the dependence on particle size cancels out and that the drift distance is therefore independent of this quantity. We find that this approximation is roughly accurate for particles in the Epstein regime with $\tau_s$ close to or greater than 1. 

Due to this effect, we consider the ``drift ice line" to be the location for which a $\tau_s = 1$ particle starting at the classical ice line location drifts and then desorbs for situations in which we know the disk surface density. If we do not know the disk surface density we use a large sized particle such that it desorbs at the maximum drift radius for a range of disk parameters.  We numerically solve for the drift distance without making any further approximations in all further discussions of drift. The existence of a maximum drift distance found in our self-consistent solution allows us to predict the circumstances in which drift affects the ice line locations. 

We find that the radial location of the classical ice line strongly impacts the amount in which drift will play a role. In particular, we find that the process of drift becomes important when the classical ice line is located where larger particles have a stopping time close to 1 as $\tau_s =1$ particles drift the fastest and larger particles take longer to desorb (see Figure \ref{stopping_disk}). The size of the $\tau_s =1$ particle depends on the overall surface density profile of the disk. Drift will therefore matter the most in determining the true ice line location at the disk radial location where the particles with $\tau_s = 1$ reach a maximum size. If the classical ice line of a particular molecule is located near this point then drift will affect the location of the ice line radius pushing it inwards. This is shown in Figure \ref{stopping_disk} through a consideration of our derived surface density profile for TW Hya and the standard profile used for the minimum mass solar nebula ($\Sigma \propto 1700 \text{ g cm}^{-2}\times r^{-3/2}$) \citep{2011ApJ...743L..16O,2010AREPS..38..493C}.

\begin{figure}\label{tauone}
\epsscale{1.1}
\plotone{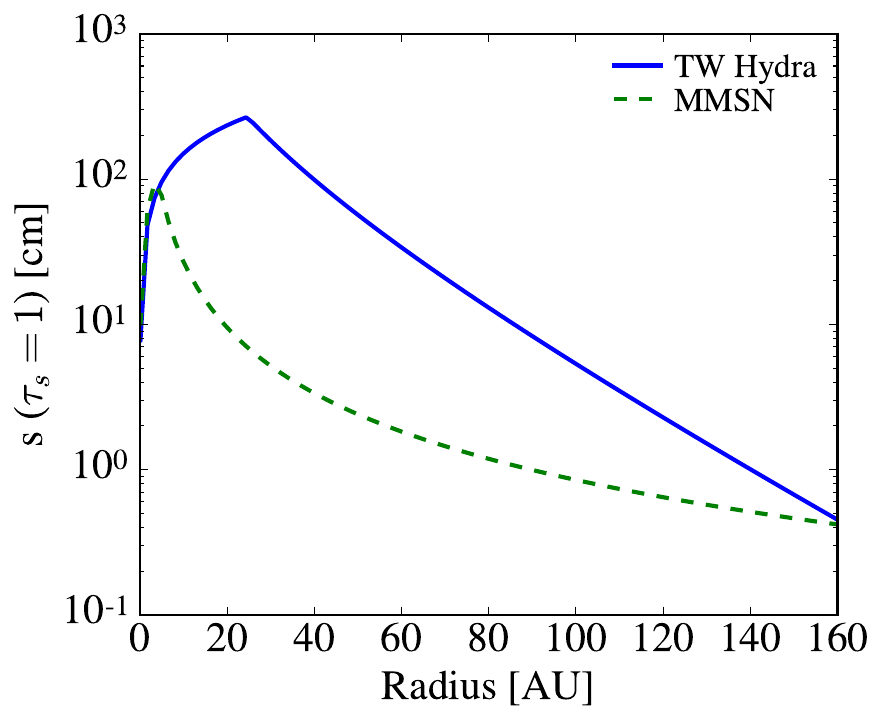}
\caption{More massive disks reach a peak in the size of $\tau_s = 1$ particles at radii further from their central star as is the case for TW Hya as compared to the minimum mass solar nebula. Shown here is the size of $\tau_s =1$ particles in the disk as a function of radius for two different surface density profiles: our TW Hya surface density profile (blue, solid) and the commonly used minimum mass solar nebula surface density profile (green, dashed). The $\tau_s = 1$ particles are the largest particles that are still well-coupled to the gas. Drift affects the ice line locations the most when the classical ice lines occur close to the peak in these plots.}
\end{figure}\label{stopping_disk}

\subsection{Application to TW Hya}\label{TW_app}
Given our derived total surface density profile for TW Hya, the last parameter needed to derive the CO ice line location is the CO abundance. We use the measured CO surface density from Schwarz et al. 2016 in conjunction with our derived total surface density profile to uncover an approximate CO abundance of 10$^{-7}$ n$_\text{H}$ from $\sim$ $10-60$ AU. This value is consistent with the average upper limit of 10$^{-6}$ n$_\text{H}$ found in \citet{2016ApJ...823...91S} which was derived by comparing their CO surface density to the surface density profile derived from HD observations of the warm gas \citep{2013Natur.493..644B}. 

This reduction of 3 orders of magnitude from the measured abundance in the ISM (10$^{-4}$ n$_\text{H}$) demonstrates that our model requires a global depletion of volatile carbon in TW Hya. \citet{2015ApJ...799..204C} infer a similar level of depletion using observations of CO and modeling by \citet{2015ApJ...807L..32D} supports this conclusion. 

We can now calculate both the classical and maximum drift CO ice lines. For our calculations we use a binding energy of $E_i/k \sim 850$ K \citep{1996ApJ...467..684A} \footnote{We note that observations discussed in \citet{2016ApJ...823...91S} indicate a CO binding energy closer to $E_i/k \sim 960$ K for TW Hya. Using this binding energy we determine that the classical CO ice line is located at $\sim$ 23 AU and the drift ice line is located at $\sim$ 10 AU. These derived values are also reasonably consistent with the observed ice lines of TW Hya (see text).}. For TW Hya, we derive a classical CO ice line of $\sim$ 30 AU and a drift ice line of $\sim$ 15 AU. 

We find that our derived classical ice line is in agreement with the ice line determined observationally via the N$_2$H$^+$ ion of $\sim$ 30 AU \citep{2013Sci...341..630Q}. We further find that our derived drift ice line is in good agreement with the CO ice line as derived through the C$^{18}$O $J=3-2$ line where a decrement of CO emission is observed until $\sim$ 10 AU \citep{2016ApJ...819L...7N}. Given the discrepancy between the ice line as determined by N$_2$H$^+$ and C$^{18}$O and the corresponding agreement we find with our theoretical model, we note that the C$^{18}$O $J=3-2$ line may be the most sensitive probe of a disk's midplane ice line, particularly when particle drift effectively smears the ice line out over a relatively large radial scale. This could be because the presence of N$_2$H$^+$ requires a lack of CO gas in the disk and may thus require that all particle sizes have frozen out (i.e. the classical ice line). 

We can now put together our complete model for the dust and ice lines in TW Hya as shown in Figure \ref{TW}. 

\begin{figure}
\epsscale{1.2}
\plotone{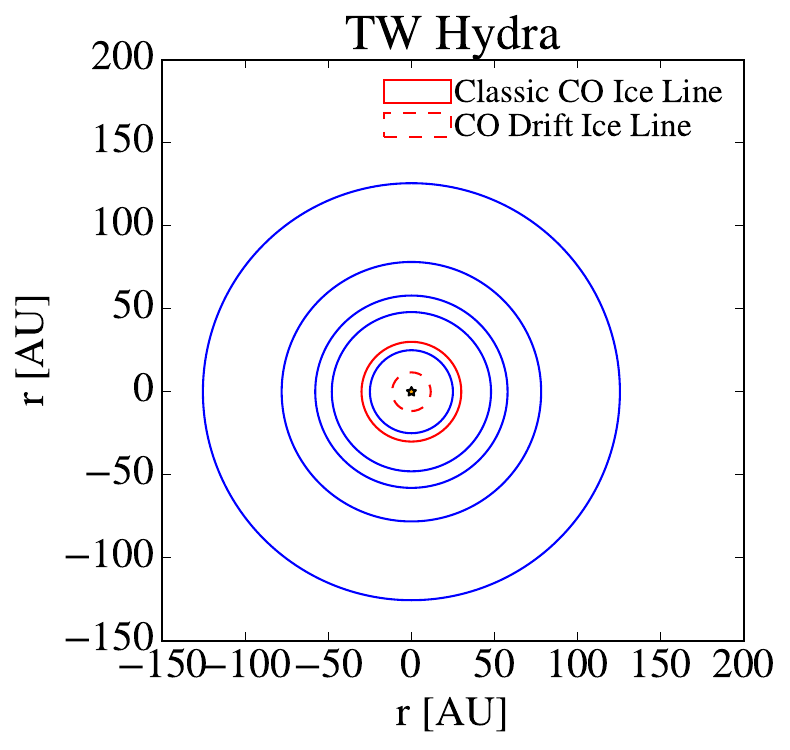}
\caption{A model of the dust and ice lines in TW Hya. The blue lines are the dust lines solved by assuming that the drift timescale is equal to the age of the system. The blue lines adequately reproduce the observed disk radial scale of TW Hya at various wavelengths. The solid red line is the classical CO ice line solved by balancing the adsorption and desorption flux onto a grain. This line is in agreement with the observed CO ice line of $\sim$ 30 AU using N$_2$H$^+$ \citep{2013Sci...341..630Q}. The dashed red line is the CO drift ice line for a $\tau_s =1$ particle at a radius of $\sim$ 15 AU which we find to be in close agreement with the ice line derived from C$^{18}$O measurements in \citet{2016ApJ...819L...7N}, suggesting that C$^{18}$O is a sensitive probe of the CO drift ice line.}
\end{figure}\label{TW}

\section{Observational Diagnostics}\label{tests}
We now propose three different observational tests of our model framework beyond the application to TW Hya. The first test is whether or not other disks have dust lines that can be well fit by a normalized form of the physically motivated surface density profile in Equation \ref{sigma_0} for disks where $r_c$ is constrained by CO observations. This is a particularly powerful test of concept if we find dust lines in other disks that are located in the power-law region of the surface density profile interior to the exponential fall-off. The second test is whether we can derive a surface density profile from the disk ice lines that is consistent with the surface density profile derived from disk dust lines. The third test is whether or not the dust and ice lines scale oppositely, as a function of surface density, across a large sample of disks.

Our second and third tests rely on an understanding of both classical (i.e. classical regime) and drift (i.e. drift dominated regime) ice lines as described in Section \ref{ice}. The particular assumptions and/or prior knowledge of a disk will inform which of the two regimes is relevant as discussed in Section \ref{driftvclassic}.

\subsection{Test 1: Surface Density from Disk Dust Lines}
The first test of our model is whether or not other disks have dust lines that can be well fit by a normalized form of the physically motivated surface density profile in Equation \ref{sigma_0}. In the era of ALMA, multiple disks will have high resolution data at several wavelengths -- the ideal data set for this fundamental test. Following the technique described in Section \ref{dust} we can convert these dust lines into a surface density measurement and see if these measurements scale with the observationally derived surface density profiles.

In the case of TW Hya, the disk dust lines were near the exponential fall-off point, which unexpectedly (due to the steep slope of the derived power law) matched previous surface density models with the application of a normalization factor. For other disks there is no reason to think that the disk dust lines will necessarily fall in the same exponential fall-off region of the disk surface density profile. If a disk has a critical radius further from the star then it is increasingly likely that we will be able to detect dust lines interior to this point. 

This test is thus a powerful test of concept that will allow us to see whether other disk sizes can be well fit through considering drift as the primary driver of disk dust lines and using these derived surface density values to normalize previous observationally fit disk surface density profiles. 

\begin{itemize}
\item \textbf{Test 1 Observational Requirements:} This test can be best carried out on disks with CO observations, such that there is a derived critical radius, and observations of dust emission at several wavelengths such that the disk dust lines are known. 
\end{itemize}

\subsection{Ice Line Regimes}\label{driftvclassic}
It is now helpful for us to distinguish between the drift and classical ice lines from observations of CO or other ice line tracers. Here we provide several rules of thumb to aid in the use of ice lines as useful diagnostics. However, we note that these aids should be used initially and then verified for self-consistent results.

Our proposed tests using these ice lines differ as the classical ice line depends on the CO abundance while the drift ice line does not. This makes the drift ice line particularly useful as it relies on fewer assumptions while the classical ice line retains its usefulness when considered across a sample of disks. 

As mentioned in Section \ref{tests}, there are two regimes that dictate when the drift or classical ice lines are relevant: the drift-dominated regime and the classical regime. In the drift-dominated regime the drift ice line is interior to the classical ice line and thus there should be only be CO freeze out exterior to this point with complete freeze-out occurring exterior to the classical ice line, making the drift ice line the most interior detectable ice line in the disk. In the classical regime particle drift does not happen quickly enough for particles to cross the ice line without desorbing and the classical ice line is observed. 

To interpret the surface density from the disk ice line (see Section \ref{test2}) we need a priori knowledge of the relevant ice line regime in the disk. Without knowledge of the disk surface density profile we can roughly determine the correct regime through an analysis of the ice line dependence on elemental abundance and disk temperature as well as a comparison of the ice line location with respect to the disk's critical radius. 

One simple metric in determining the ice line regime is the knowledge that the ice line location is preferentially dominated by drift for disks that have large particles with a stopping time of unity near the ice line location. As seen in Section \ref{driftice}, particles can drift further without desorbing where the $\tau_s \approx1$ particle size is large. This size increases at a given radius with increased disk critical radii and surface density as demonstrated in Figure \ref{tone} where we calculate the $\tau_s = 1$ particle size at 30 AU for a disk with the same temperature profile as TW Hya for a range of surface density normalization factors and disk critical radii. We find that an ice line observed interior to the critical radius is likely dominated by drift for massive disks with large critical radii. 

\begin{figure}
\epsscale{1.2}
\plotone{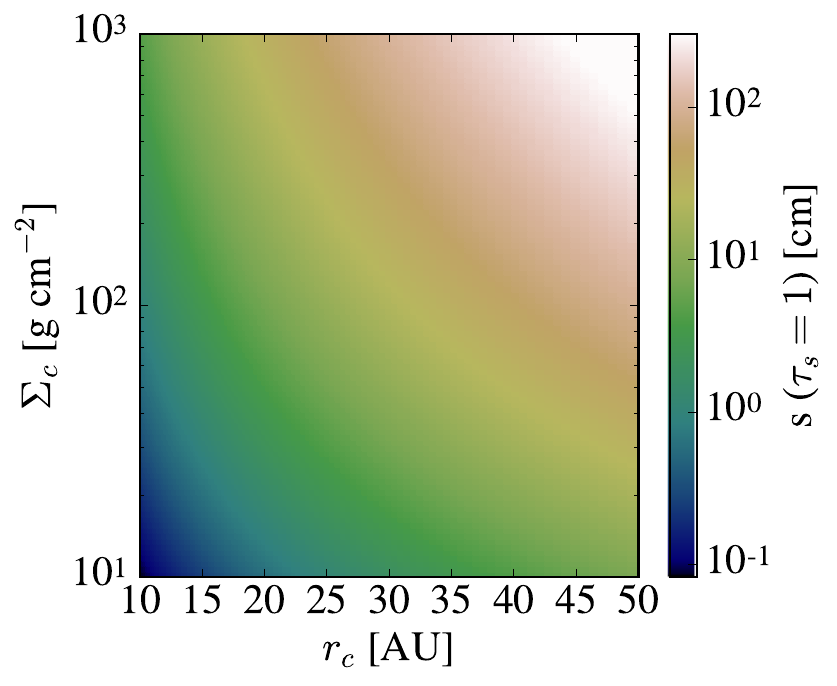}
\caption{The $\tau_s = 1$ particle size for a disk using our derived temperature profile for TW Hya (see Equation \ref{canon} with $T_0 = 82$ K) at a radius of 30 AU (the observed classical ice line radius). We vary the disk critical radius and the surface density normalization and find that the ice line is likely to be dominated by drift for disks with large critical radius and high surface density normalization. }
\end{figure}\label{tone}

Another parameter that determines the ice line regime is the amount of radiation that the disk receives from its host star (i.e. the stellar luminosity). The importance of this parameter is shown in Figure \ref{trends} for generally assumed molecular abundances \citep[$\text{n}_\text{CO} =1.5\times10^{-4}\;\text{n}_\text{H}$, $\text{n}_{\text{CO}_2}=0.3\times10^{-4}\;\text{n}_\text{H}$, $\text{n}_{\text{H}_2\text{O}} =0.9\times10^{-4}\;\text{n}_\text{H}$;][]{2006A&A...453L..47P}.

\begin{figure*}
\epsscale{1.1}
\plotone{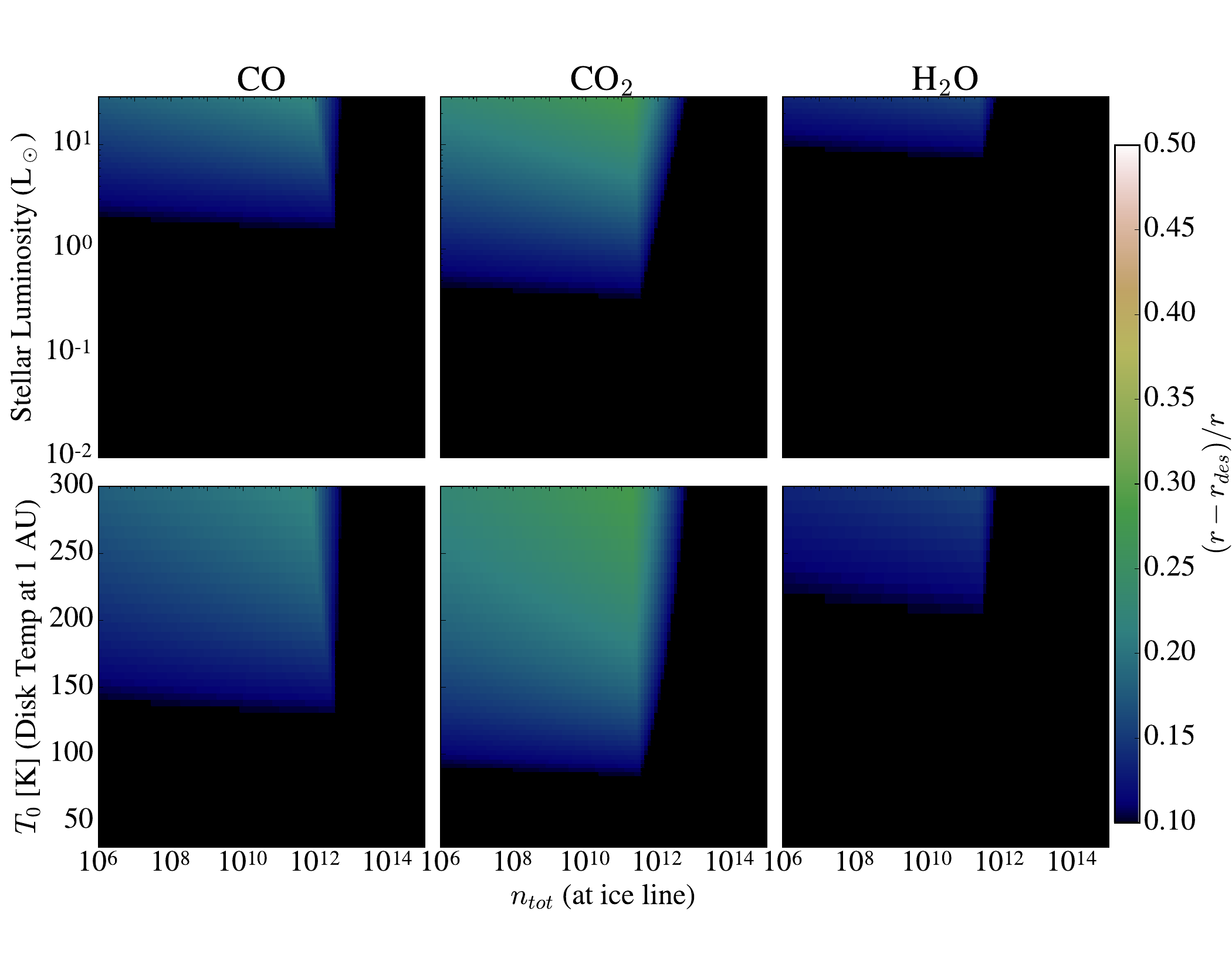}
\caption{The fractional difference between the classically derived ice line $r$ and the drift ice line $r_{des}$ as a function of stellar luminosity and disk density (top) as well as disk density and temperature (bottom) for the following molecular abundances: $\text{n}_\text{CO} =1.5\times10^{-4}\;\text{n}_\text{H}$, $\text{n}_{\text{CO}_2}=0.3\times10^{-4}\;\text{n}_\text{H}$, $\text{n}_{\text{H}_2\text{O}} =0.9\times10^{-4}\;\text{n}_\text{H}$ \citep{2006A&A...453L..47P}. Drift is most important for CO$_2$ and generally increases at moderate densities and high temperatures. The y-axis label $T_0$ refers to the temperature normalization for Equation \ref{canon} and we can convert from this temperature normalization to stellar luminosity using Equation \ref{T0}. For stellar luminosities below $10^{-2}$ L$_{\odot}$ drift does not play a role in determining the ice line locations.}
\end{figure*}\label{trends}

We find that we are in the drift dominated regime for disks with moderate densities and high temperatures across all three molecular species. At very high disk densities we find that the drift ice line is irrelevant and only the classical ice line can be observed, as a large increase in disk density moves the classical ice line inwards such that the disk temperature at the ice line is higher and the desorption timescale is shorter (i.e. the increased temperature near the star overwhelms the effect of the increase in disk density, see Equation \ref{de}). As the disk density is not known a priori we posit that the drift ice line is likely relevant for disks with high stellar irradiation.

We further find that, given a molecular depletion as in the case of our derived disk parameters for TW Hya and may be true for the majority of disks, drift sets the true ice line location across a wide range of parameter space as shown in Figure \ref{xco_tempdens}. Drift is more important when the CO abundance is smaller because the decrease in CO abundance moves the classical ice line further from the star while the drift ice line remains unchanged with the change in abundance. Therefore, we find that disks depleted in CO should have ice lines that are determined by drift. This is further demonstrated in Figure \ref{COfrac} where, across a wide range of molecular CO abundances, the drift ice line is constant and interior to the classical ice line given our derived disk parameters for TW Hya.

\begin{figure}
\epsscale{1.1}
\plotone{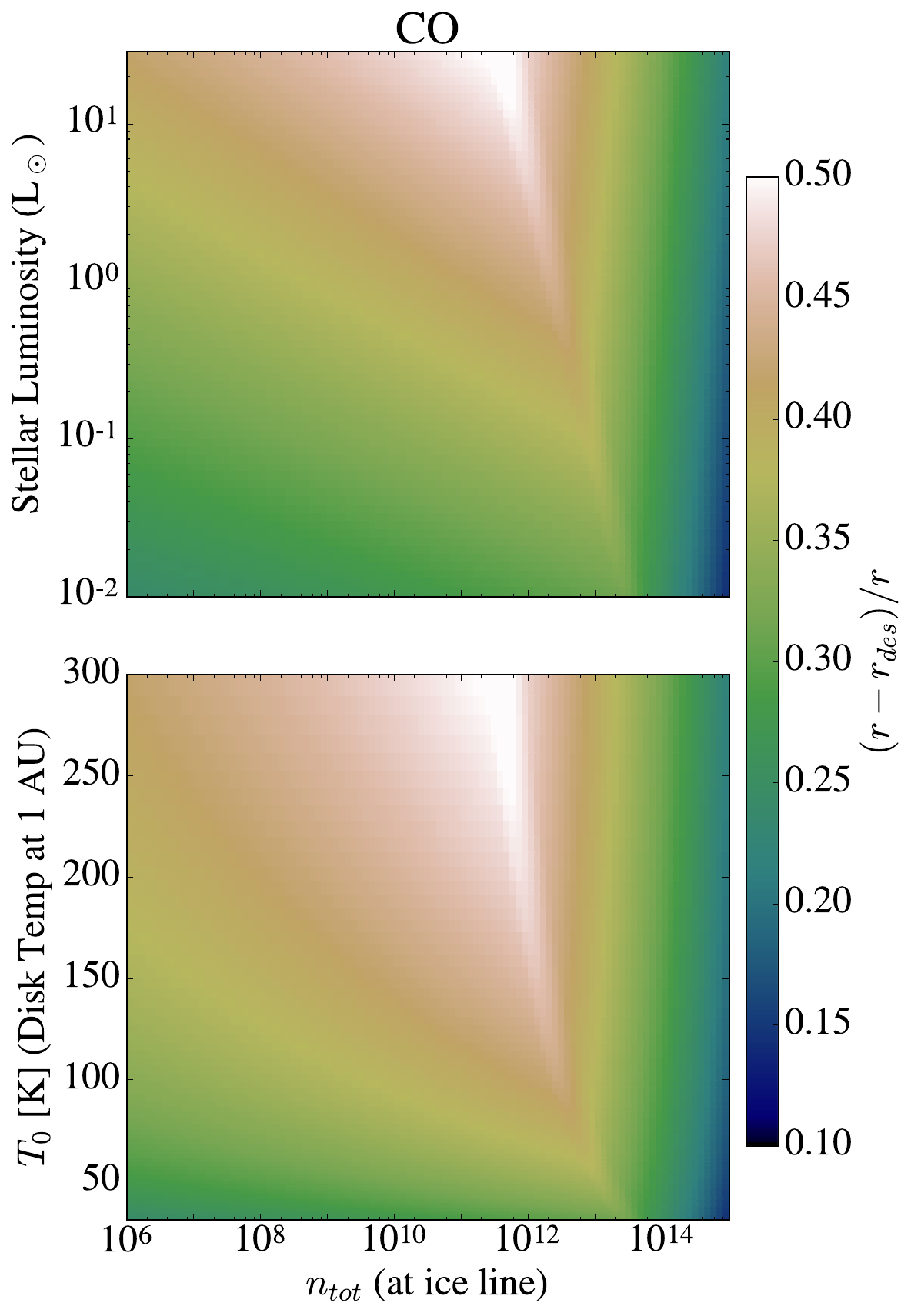}
\caption{The fractional difference between the classically derived ice line $r$ and the drift ice line $r_{des}$ as a function of stellar luminosity and disk density for an $\text{n}_{\text{CO}}$ of 10$^{-7}$ n$_\text{H}$. We find that drift plays a role in determining the extent of the ice line location across the full range of our parameter space. }
\end{figure}\label{xco_tempdens}

\begin{figure}
\epsscale{1.1}
\plotone{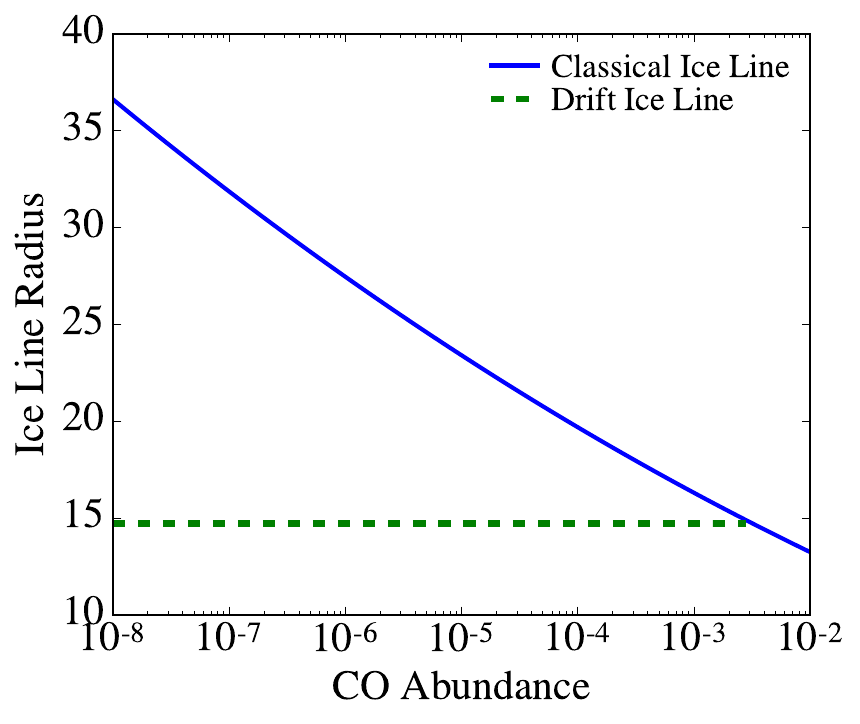}
\caption{The classical ice line location (blue) and drift ice line location (green, dashed) as a function of CO abundance for our derived disk surface density and temperature profile for TW Hya. The drift ice line location is constant and interior to the classical ice line radius for a wide range of CO abundances.}
\end{figure}\label{COfrac}

It is therefore a relatively safe assumption to consider the CO ice line location to be determined by drift for disks that are hot, depleted in CO, and/or have an observed ice line interior to a large critical radius and appear to be relatively massive. Disks for which the opposite is true are in the classical regime and their ice lines should be theoretically treated as classical ice lines. 

\subsection{Test 2: Surface Density from Disk Ice Lines}\label{test2}
The second test of our model is whether or not the surface density profile derived from a consideration of the disk ice line matches the surface density profile derived from the disk dust lines. We now discuss how empirical knowledge of a disk's molecular ice line can be used to independently determine the total surface density profile of a disk. This can be done as both the drift and classical ice lines depend on the disk surface density. 

\subsubsection{Drift Ice Line Surface Density Determination}
If the disk in question is in the drift dominated ice line regime (see Section \ref{driftvclassic}), solving for the disk surface density profile simply involves setting Equation \ref{rdot} for $\tau_s = 1$ (where particle drift reaches a maximum) equal to Equation \ref{desd} and solving for surface density given an assumed characteristic particle size. This surface density measurement can then be used to normalize the total disk surface density profile, thus giving us an independent estimate that can be compared to the profile derived via the disk dust lines. 

While this method requires the assumption of a characteristic particle size, we note that a wide range of particle sizes will result in the same drift ice line location (see Figure \ref{drift_distance}) and thus, the derived surface density profile will not be sensitively affected by this value. For computational purposes we use a particle size of 1 m such that its stopping time exceeds unity for a wide range of disk parameter space. As is the case for all particles close to or larger than a stopping time of unity, these particles should desorb after drifting to the maximum drift ice line.

Consideration of the drift ice line in determining disk surface density has the advantage of not needing a measure of the molecular abundance in the disk. Regardless of CO abundance this location will be constant as it is dependent on disk surface density and not the CO surface density. Thus, a sensitive probe of the CO ice line (i.e. the C$^{18}$O $J=3-2$ emission) should be able to detect the uptick in CO emission past this point and provide a probe of disk surface density from this measurement alone without further assumption. 

\subsubsection{Classical Ice Line Surface Density Determination}

If the disk in question is in the classical ice line regime (see Section \ref{driftvclassic}), the surface density can be derived by balancing Equation \ref{ad} and Equation \ref{de}. Unlike the drift dominated regime, this calculation requires an assumed molecular abundance. While these uncertainties may diminish the robustness of this test, we can make the simplifying assumption that molecular abundances are roughly constant across a single stellar type. This could be a reasonable assumption if molecular abundances are primarily shifted from ISM values via photochemical processes. If we accept this assumption as true, we can determine the disk surface density profile for disks of a stellar type for which one member has a relatively well determined molecular abundance. We note that, for this method, a factor of 2 change in the ice line location will change the derived surface density by a factor of 3.

The relative behavior of the classical ice line across many disks will also improve this test and may indeed allow the potential for deriving the CO abundance in conjunction with Test 1.

\begin{itemize}
\item \textbf{Test 2 Observational Requirements:} This test can be carried out on disks with accurate ice line measurements (either via C$^{18}$O or N$_2$H$^+$). To carry out this test more precisely it is also useful to have observationally derived CO surface density profiles for these disks.
\end{itemize}

\subsection{Test 3: Disk Dust and Ice Line Scalings}\label{test3}
The third test is whether or not the dust and ice lines scale oppositely across a large sample of disks. Our model predicts that this will happen as disks with dust lines at larger radial scales should have ice lines located at shorter radii across a particular disk temperature profile and molecular abundance. This arises as larger particles become well coupled to the gas as disk surface density increases, thus causing a given particle size to have a dust line located further out in the disk. Conversely, increased surface density increases the adsorption flux onto a grain such that the freezing temperature is hotter and the ice line is moved closer to the star. 

To clearly see how the dust and ice lines in a disk scale as a function of surface density alone, we vary the disk surface density for a set disk temperature profile. Using the empirical evidence that the Earth is not formed of water ice, we consider the case of a passively irradiated disk that is normalized such that the H$_2$O ice line is outside of 1 AU as shown in Figure \ref{tempG} as our fiducial temperature profile. We take this temperature profile to be roughly representative of a young sun like star. This profile follows Equation \ref{canon} with a derived $T_0 \approx 210$ K. 

This profile is derived through the use of an assumed canonical H$_2$O abundance of 0.9$\times10^{-4}$ n$_\text{H}$ \citep{2006A&A...453L..47P}. While this assumption is sufficient for our illustrative example, we again note that disk molecular abundances may be poorly constrained. However, let us naively assume that molecular abundances vary from ISM values as a function of photochemistry such that they are constant across a particular stellar type and thus a particular temperature profile. Thus, using this temperature profile we can derive the trends that we expect to see as we vary the disk surface density. 

\begin{figure}
\epsscale{1.15}
\plotone{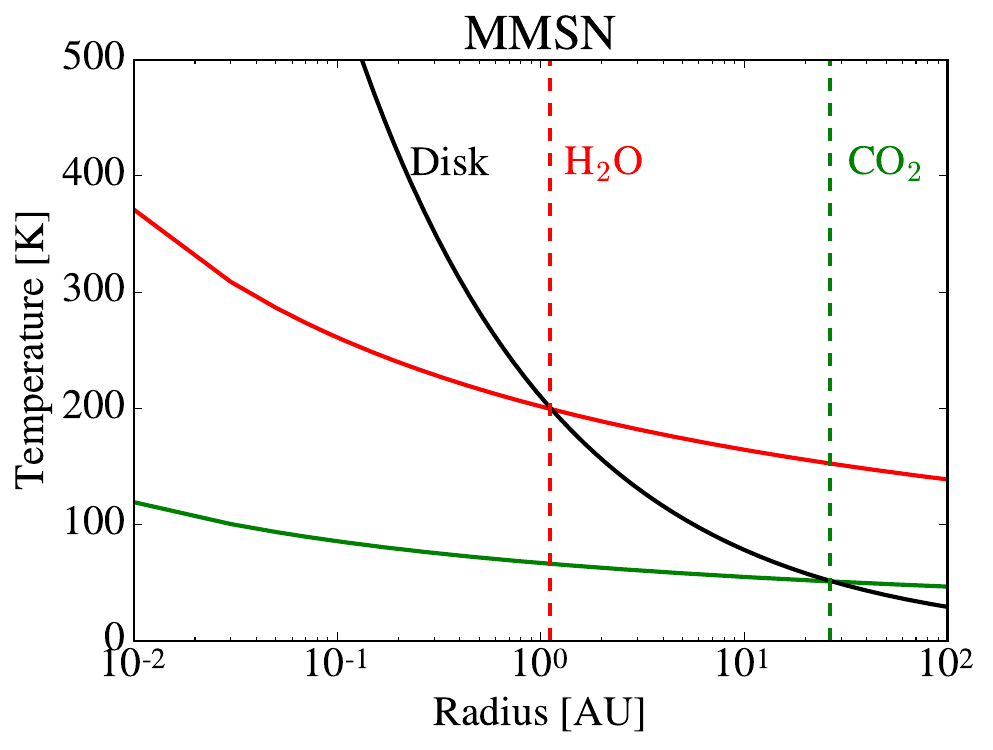}
\caption{The H$_2$O (red) and CO$_2$ (green) freezing temperature (calculated by self-consistently balancing Equations \ref{ad} and \ref{de}) as a function of radius for the minimum mass solar nebula (MMSN). The black line is the minimum temperature profile that places the H$_2$O ice line outside of 1 AU. Classical snow lines occur where the freezing temperature and the disk temperature are equal (dashed lines). }
\end{figure}\label{tempG}

We find that for all three of our molecular species: CO, CO$_2$, and H$_2$O, the radial extent of the disk increases with increasing surface density while the ice line location decreases as expected (see Figure \ref{CO_trend}). For the CO and H$_2$O ice lines, the drift distance past the classical ice line is negligible while for the CO$_2$ ice line drift plays a small role that decreases in importance with increasing surface density. We note that while the importance of the drift ice line will vary with disk parameters, the general trend should hold.

This trend remains true across a range of molecular abundances as long as they are held constant across a sample of disks. Here we have assumed the following abundance values for our molecular species: $\text{n}_\text{CO} =1.5\times10^{-4}\;\text{n}_\text{H}$, $\text{n}_{\text{CO}_2}=0.3\times10^{-4}\;\text{n}_\text{H}$, $\text{n}_{\text{H}_2\text{O}} =0.9\times10^{-4}\;\text{n}_\text{H}$ \citep{2006A&A...453L..47P}. We note that lower molecular abundances result in lower freezing temperatures (see Section \ref{classical}) which, depending on the temperature profile of the disk, pushes the classical ice lines further from the star with the reverse being true for higher molecular abundances.

\begin{figure}
\epsscale{1.14}
\plotone{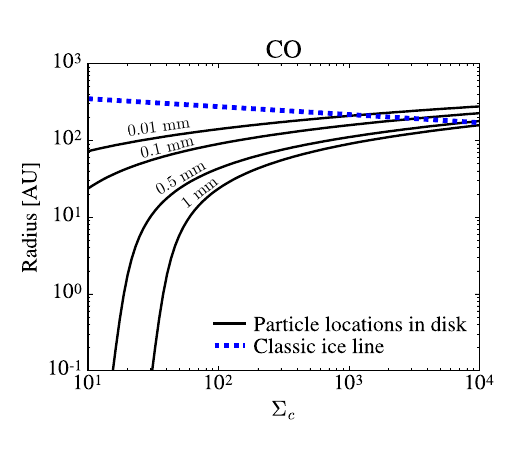}
\epsscale{1.11}
\plotone{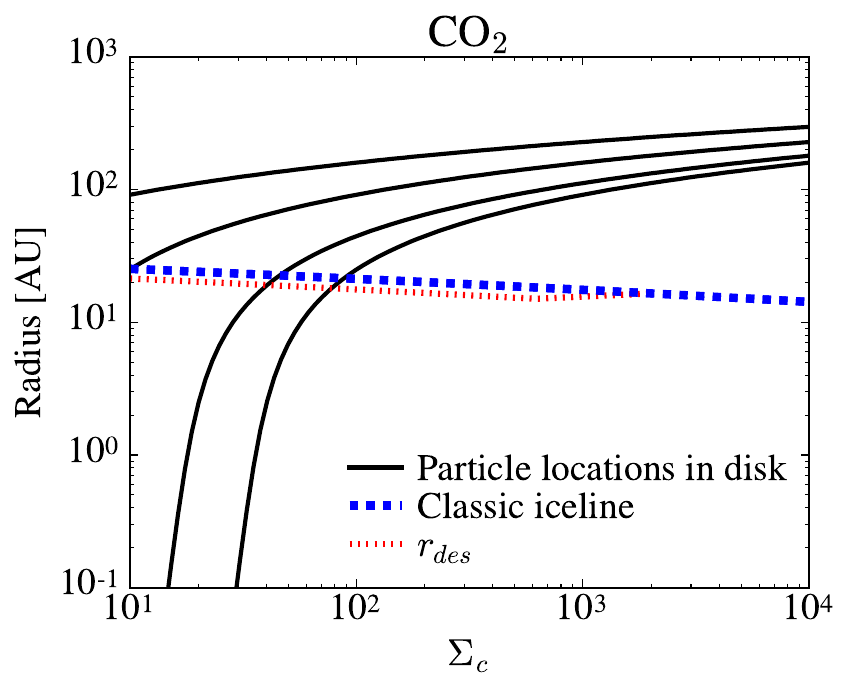}
\plotone{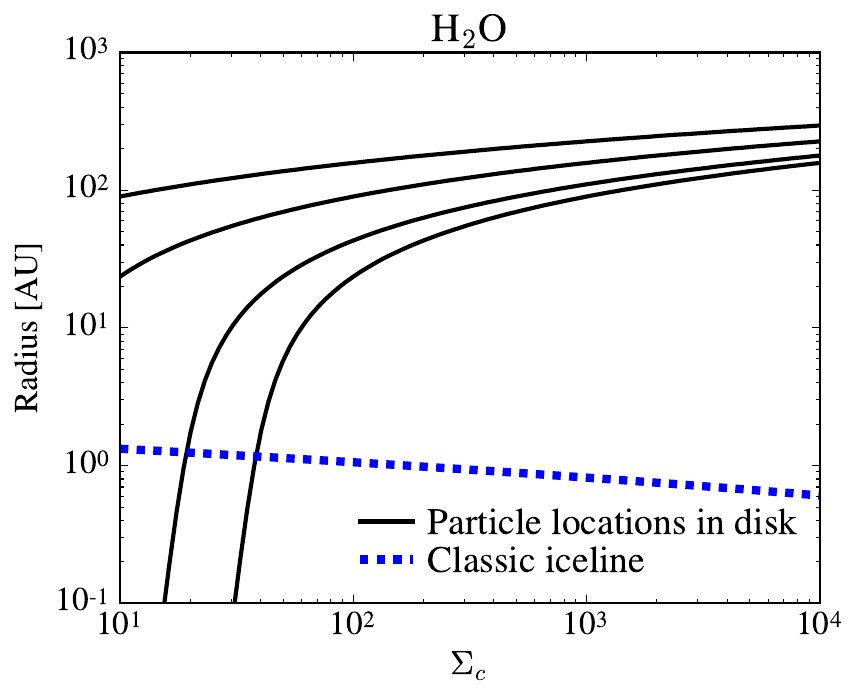}
\caption{The disk radial scale for different particle sizes (black, solid), the classically derived ice line locations (blue, dashed) and the drift ice line (red, dotted) for CO (top; with particle sizes labeled), CO$_2$ (middle), and H$_2$O (bottom) as a function of surface density. The dust line locations increase dramatically with increased surface density while the ice line location decreases.}
\end{figure}\label{CO_trend}

This trend is therefore a key observational diagnostic that will be confirmed if, across a constant molecular abundance, disks with dust lines at larger radii have ice lines closer to their star. We also note that, if drift determines the ice line location (see Section \ref{driftvclassic}) we can observe these trends without the need for an assumed molecular abundance. 

\begin{itemize}
\item \textbf{Test 3 Observational Requirements:} To best perform this test there needs to exist a significant sample of disks with observations of both the dust and ice lines. 
\end{itemize}

\section{Summary \& Discussion}\label{discussion}
We propose a novel method to derive disk surface density through the consideration of disk dust and ice lines. To derive this method we adopt an agnostic point of view in regards to disk surface density which we do in particular response to the uncertainties that accompany typical observational tracers. This method relies on the assumption that, at late stages of evolution, the growth timescale, drift timescale, and the lifetime of the disk are all equal for the dominant particles at a dust line location ($s=\lambda_{\text{obs}}$). 

While other work finds that these timescales are equal they are often unable to match the disk surface density. We therefore make the assumption that these timescales are equal and use this to determine the surface density profile without evoking other observational tracers. These assumptions allow us to self-consistently derive a disk surface density profile as well as a dust-to-gas ratio in the outer disk without the further assumption of a given molecular abundance. 

We apply our modeling technique to our fiducial disk TW Hya. We find that our derived surface density profile and dust-to-gas ratio are consistent with the lower limits found through measurements of the HD gas in the disk \citep{2013Natur.493..644B,2016ApJ...823...91S}. Using our derived surface density profile we uncover a theoretical estimate of the disk accretion rate that is more closely aligned with the measured accretion onto TW Hya. We further find that our theoretical classical and drift ice lines have clear observational analogues where the classical ice line is predictive of the ice line found via N$_2$H$^+$ emission \citep{2013Sci...341..630Q} and the drift ice line is probed by the C$^{18}$O emission \citep{2016ApJ...819L...7N}. We conclude that the ice line derived through observations of C$^{18}$O emission may be more sensitive to the extent of the drift ice line in the disk and thus may be the best observational method to test our model's assumptions. We note here that our method highlights the likelihood of detecting multiple ice lines for a given molecular species as there will be relatively large regions in the disk where some of the material has desorbed and some of the material has not. 

Furthermore, if the banded structure in the ALMA images of HL Tau and TW Hydra reflect substantial variations in surface density then our model should still hold. However, ice lines may be preferentially found in less dense bands in the disk where drift slows. 

We next consider a large range of theoretical disk parameter space and uncover three observational tests of our model. The first test is whether or not the dust lines of other disks, once converted to surface densities, can be matched with a previously derived normalized surface density profile for the disk. For TW Hya, the disk dust lines fall near the exponential cut off region of the disk but there is no reason that this needs to be true across many objects and thus provides a powerful test of the model. In the age of ALMA, dust lines can be determined observationally for a significant sample to protoplanetary disks which will provide the ideal observations necessary for this test. 

The second test is whether we can derive a surface density profile from the disk ice lines that matches the surface density profile derived via disk dust lines. The third and final test is whether or not disk dust and ice lines scale oppositely, as a function of surface density, across a large sample of disks. Disk ice lines have been observed for an increasing number of disks such that the last two tests could be carried out in the near future with the aid of facilities such as ALMA.

\section{Acknowledgements}
This work was done as a part of the Kavli Summer Program in Astrophysics: Exoplanetary Atmospheres 2016. We would like to thank Jonathan Fortney and Pascale Garaud for providing us with the opportunity to work together on this project and for organizing a wonderful workshop. We thank James Owen, Michael Rosenthal, Catherine Espaillat and Laura P\'{e}rez for their useful comments and insightful discussion. This material is based upon work supported by the National Science Foundation Graduate Research Fellowship under Grant DGE1339067. RMC acknowledges support from NSF grant number AST-1555385.

\bibliography{refs}
\bibliographystyle{aasjournal}

\end{document}